\documentclass[12pt]{article}
\pdfoutput=1

\usepackage{bbold}
\usepackage{amssymb}
\usepackage{graphics}
\usepackage{graphicx}
\usepackage{lscape}
\usepackage{multirow}
\usepackage{amssymb,amsmath,amsthm}
\usepackage{amstext}
\usepackage{amscd}
\usepackage{epsfig}
\usepackage[mathscr]{eucal}
\usepackage{times}
\usepackage{srcltx}
\usepackage{shadow}
\usepackage{hyperref}

\def\wt#1{\widetilde{#1}}

\def\vphi{\varphi}

\def\be{\begin{equation}}
\def\ee{\end{equation}}
\def\beq{\begin{equation}}
\def\eeq{\end{equation}}
\def\bea{\begin{eqnarray}}
\def\eea{\end{eqnarray}} 
\def\beqa{\begin{equation}\begin{array}{l}}
\def\eeqa{\end{array}\end{equation}}

\def\eqn#1{(\ref{#1})}

\def\eqref#1{eq.~(\ref{eq:#1})}

\def\vphi{\varphi}

\def\pa{\partial}

\def\pa{\partial}

\def\nn{\nonumber}

\newcommand{\Rho}{{\mbox{\sf P}}}

\def\sideremark#1{\ifvmode\leavevmode\fi\vadjust{\vbox to0pt{\vss
 \hbox to 0pt{\hskip\hsize\hskip1em
 \vbox{\hsize3cm\tiny\raggedright\pretolerance10000
  \noindent #1\hfill}\hss}\vbox to8pt{\vfil}\vss}}}

\begin{document}

\thispagestyle{empty}

\vspace{.8cm}
\setcounter{footnote}{0}
\begin{center}
{\Large{\em 
LOCAL~UNIT~INVARIANCE,~BACK-REACTING\\[1mm] TRACTORS AND THE COSMOLOGICAL\\[4mm] CONSTANT PROBLEM}
    }\\[10mm]

{\sc R. Bonezzi$^{\mathfrak B}$, O. Corradini$^{\mathfrak C}$
and A. Waldron$^{\mathfrak W}$
\\[6mm]}

{\em\small  
${}^{{\mathfrak B}, {\mathfrak C}}$ Dipartimento di Fisica, Universit\`a
  di Bologna and INFN, Sezione di Bologna\\ via Irnerio 46, I-40126 Bologna, Italy
\\ {\tt bonezzi,corradini@bo.infn.it}}\\[5mm]
{\em\small  
${}^{\mathfrak W}$ Department of Mathematics,
University of California, Davis, CA 95616,
USA\\ {\tt wally@math.ucdavis.edu}}\\[5mm]

\bigskip

\bigskip
{\sc Abstract}\\
\end{center}

{\small
\begin{quote}

When physics is expressed in a way that is independent of local choices of unit systems,
Riemannian geometry is replaced by conformal geometry. Moreover masses become geometric,
appearing as Weyl weights of tractors (conformal multiplets of fields necessary to keep local unit invariance
manifest). The relationship between these weights and masses is through the scalar curvature. As a consequence
mass terms are spacetime dependent for off-shell gravitational backgrounds, but happily constant for physical,
Einstein manifolds. Unfortunately this introduces a naturalness problem because the scalar curvature is proportional 
to the cosmological constant. By writing down tractor stress tensors (multiplets built from the standard
stress tensor and its first and second derivatives), we show how back-reaction solves this naturalness problem.
We also show that classical back-reaction generates an interesting potential for scalar fields.
We speculate that a proper description of how physical systems couple to scale, could improve our understanding 
of naturalness problems caused by the disparity between the particle physics and observed, cosmological constants. 
We further give some ideas how an ambient description of tractor calculus could lead to a Ricci-flat/CFT
correspondence which generalizes the AdS side of Maldacena's duality
to a Ricci-flat space of one higher dimension.



\bigskip

\end{quote}
}

\newpage





\section{Introduction and Summary}
\label{Introduction}

One of the main puzzles of modern physics is undoubtedly the
cosmological constant problem. It can be
viewed as a naturalness problem, as there are no known dynamical
mechanisms that allow relaxation of the huge particle physics vacuum energy
density into the observed tiny acceleration of the
universe~\cite{Weinberg:1988cp} (for more recent surveys see~\cite{Carroll:2000fy}). 
To date the only way to reconcile these two quantities, within local
four-dimensional field theory, is by a very unnatural
fine-tuning. (Non-locally modified theories of gravity might
  account for an effective scale-dependent gravitational
  constant that works as a high-pass filter, becoming tiny in the far
  infrared limit~\cite{ArkaniHamed:2002fu}.) 
From a general point of view this problem can be seen
as a problem of coupling between particle physics (that yields the
vacuum energy density) and gravity, where the vacuum energy density
acts as a source: particle physics phenomena take place at
microscopic scales whereas the cosmological constant displays its
effects only at enormous scales, since the cosmological constant does not red-shift with the
expansion of the Universe (as opposed to matter gravitational
fluids). In essence the cosmological constant is generated by ultraviolet
features of the theory, through vacuum diagrams, but manifests itself {\em only} in the far
infrared, and only through gravitational couplings: particle physics
observables do not depend on it.     
In fact, in general, particle physics is formulated in a (flat) non-dynamical
background and therefore knows nothing about local rescalings of the metric. 

The  standard description of particle physics in terms of a flat background 
could well underly our lack of understanding of this
apparently  enormous hierarchy. In fact quantum effects (particle
creation, vacuum polarization) in de Sitter
spacetime might be the reason for the smallness of the observed Hubble
rate~\cite{Polyakov:2007mm} as they may lead to a screening of the particle
physics vacuum energy density~\cite{Antoniadis:2006wq}. Hence, perhaps 
one way to study the cosmological constant problem 
is to have a theory that manifestly incorporates the effects of local scale transformations.
Or more precisely, what may be needed is a formulation of physics in terms of conformal geometry
rather than Riemannian geometry. In that case, the flat particle physics background 
would be replaced by a conformally flat class of metrics and the information of local scale 
transformations would not be lost. To be sure, in this letter we do not purport to solve
the cosmological constant problem by taking this viewpoint, but we do
show how it allows the solution of  a naturalness problem for a
particularly interesting class of models with off-shell spacetime
dependent mass terms. (It would of course be very interesting to make
contact with the approach described in~\cite{Antoniadis:2006wq}.)

A fundamental principle of physics is that local choices of unit
systems---{\it local unit invariance}---could  
not possibly change the outcome of any physical measurement~\cite{Gover:2008sw}.
Therefore there must exist a formulation of physics that makes this symmetry
manifest. (A similar line of reasoning led Einstein to postulate a theory of gravitation
in terms of (pseudo)Riemannian geometry in order to manifest a local coordinate invariance.)
This simple idea led Weyl to a study of local metric transformations of the form~\cite{Weyl:1918ib}
\be\label{Weyl}
g_{\mu\nu}\mapsto \Omega^2(x) g_{\mu\nu}\, ,
\ee
which are by now called Weyl transformations. There exist relatively few physical models
that exhibit this symmetry, notable examples include four-dimensional Maxwell theory, conformally
improved scalars, the massless Dirac equation and Weyl-squared gravity. In general, just as general
coordinate invariant theories typically require the introduction of a metric~$g_{\mu\nu}(x)$ (roughly speaking the gauge field for local translations), local unit invariant theories require a new gauge field called the scale~$\sigma(x)$, whose
value at differing spacetime points reflects the ratio of unit systems at those points. It can also be viewed as a spacetime dependent Newton ``constant''. In the physics literature, the scale~$\sigma(x)$ is often called a dilaton 
or Weyl compensator~\cite{Zumino,Deser:1970hs}. Under local changes of unit systems, the scale transforms as
$$
\sigma\mapsto \Omega(x) \sigma\, .
$$
In mathematical terms, the symmetry~\eqn{Weyl} implies that physics can be formulated in terms of conformal geometry (the theory of conformal classes of metrics $[g_{\mu\nu}]=[\Omega^2 g_{\mu\nu}]$). It is important to note 
that the choice of gauge (local Weyl frame)
$$
\sigma=\kappa^{\frac2{d-2}}=M_{\rm Pl}^{-1}\, , 
$$
both yields the standard description of physics with a constant value
of the Newton constant (or equivalently, Planck mass) and selects a distinguished (or ``canonical'')
metric from the double equivalence class
$[g_{\mu\nu},\sigma]=[\Omega^2g_{\mu\nu},\Omega\sigma]$; in other words
the scale is precisely the field that defines the gravitational coupling.

Without a tensor calculus for rapidly constructing Weyl invariant
quantities, the above local unit invariance principle 
would not be particularly enlightening. Fortunately, such a calculus
already exists in the mathematical literature and goes under the name ``tractor calculus''~\cite{Thomas,BEG,G,GP,CG}.
It is the mathematical machinery required to replace Riemannian geometry with conformal geometry
as the underpinning of physics. A particularly appealing implication is that in a description of physics
that manifests  local unit invariance, masses are replaced by Weyl weights which measure the 
response of physical fields to changes of unit systems. In particular, mass terms become spacetime
dependent in general gravitational backgrounds, yet constant when these backgrounds are Einstein.
Unfortunately, the constant of proportionality in the relationship between standard weights and masses
is the cosmological constant~\cite{Gover:2008sw}. This yields a naturalness problem because
extremely large
numbers (weights) of order $10^{60}$ are required to compensate the smallness of the cosmological constant.
We show however
that, once back-reaction is taken into account, the problem is relaxed. 
Moreover, the mechanism we use is invisible to the standard particle physics description in flat backgrounds.

Our computations rely on new results for stress-energy tensors: the four-dimensional stress tensor, its trace, first and second derivatives plus couplings to gravity in fact form a six-dimensional, rank two, symmetric tractor multiplet. 
Using this fact, we can   write down a matter tractor stress
tensor $\frak T_{MN}$, and a gravity tractor stress tensor $\frak G_{MN}$ 
(which subsumes the standard Einstein tensor) which together give a
system of coupled tractor equations describing back-reaction. In vacuum, the
vanishing of the gravity stress tensor correctly yields (conformally)
Ricci-flat solutions, whereas the equation $\frak G^M_N -\frac1{d+2}
\delta^M_N \frak G^R_R=0$ yields (conformally) Einstein spaces.  
These computations are presented in section~\ref{TST}, while section~\ref{tractors} briefly reviews 
the conformal geometry and tractor description of physics.
Our solution to the above naturalness problem via back-reaction is given in section~\ref{TBR}.
The final section contains some more speculative ideas how the classical framework we present
can be applied to quantum field theories, which generically contain scale anomalies.
The appendix presents a rather interesting analysis of the non-linear potential terms generated by 
a classical analysis of a simple scalar field toy model.

\section{Conformal Geometry and Physics}
\label{tractors}
Tractor calculus is closely related to many developments in the physics literature
such as the gauging of spacetime algebra method applied to the
conformal group~\cite{Kaku:1977pa}, Bars' two-time formalism~\cite{Bars:1996dz}
and Boulanger's conformal tensor calculus~\cite{Boulanger:2004eh}. The
key idea of tractor calculus is that (in four dimensions),
four-vectors should be elevated to six-vectors, an idea that harks
back to Dirac's formulation of conformal systems in six
dimensions~\cite{Dirac}. These six-vectors, or in arbitrary dimensions~$d$,
$(d+2)$-multiplets, are required to transform under local scale transformations as
\be\label{tractor gauge transformation}
V^M \mapsto \Omega^w U^M{}_N V^N\, , \qquad
U^M{}_N=
\begin{pmatrix}
\Omega & 0 & 0 \\[2mm]
\Upsilon^m & \delta^m_n&0 \\[2mm]
-\frac{\Upsilon^2}{2\Omega} & -\frac{\Upsilon_n}\Omega &\frac1\Omega
\end{pmatrix}\, .
\ee
In these formul\ae
$$
\Upsilon_\mu = e_\mu{}^m \Upsilon_m =\Omega^{-1}\partial_\mu \Omega\, ,
$$
and the special gauge transformation~$U^M{}_N$ is called a {\it tractor gauge transformation}.
It defines sections of a tractor vector bundle~$V^M$. The index $M$ takes on $d+2$ values
$(+,m,-)$. It is extremely important to realize that the lower block triangular form of the tractor gauge 
transformation~$U^M{}_N$ allows the tractor vector~$V^M$ to carry a weight~$w$. This tractor weight will
describe the mass of the field~$V^M$. This leads to a general picture where geometric properties
of tractors (their weight and tensor rank) will correspond to physical properties (mass and spin)
in generally curved spaces.

For the purpose of this paper, very few elements of tractor calculus will be needed, for a general description we refer the reader to some of the original mathematics and physics papers~\cite{Gover:2008sw,Gover:2009vc,Thomas,BEG,G,GP,CG}. Perhaps the most fundamental ingredient is the 
Thomas~$D$-operator 
$$
D^M=
\begin{pmatrix}
w(d+2w-2)\\[2mm] 
(d+2w-2) {\cal D}^m\\[2mm] 
-(g^{\mu\nu}{\cal D}_\mu{\cal D}_\nu+
w\Rho)
\end{pmatrix}\, ,
$$
which maps rank~$k$, weight~$w$ tractors to weight~$w-1$, rank~$k+1$ tractors. It is built from the tractor covariant derivative
\be\nn
{\cal D}_\mu = \begin{pmatrix}\partial_\mu&-e_{\mu n} & 0 \\[1mm]
  \Rho_\mu{}^m
&\nabla_\mu{}^m{}_n&e_\mu{}^m\\[1mm]0&-\Rho_{\mu n}&\partial_\mu\end{pmatrix}\, ,
\ee
(which transforms under Weyl transformations as~${\cal D}_\mu\mapsto U{\cal D}_\mu\, U^{-1}$) and in turn
the Schouten tensor defined by the difference of Riemann and Weyl tensors via
\be\nn
R_{\mu\nu\rho\sigma}-W_{\mu\nu\rho\sigma}=
g_{\mu\rho}\Rho_{\nu\sigma}-g_{\nu\rho}\Rho_{\mu\sigma}-g_{\mu\sigma}\Rho_{\nu\rho}+g_{\nu\sigma}\Rho_{\mu\rho}\,  ,
\ee
and its trace~$\Rho=\Rho_\mu^\mu$.

With no more than the above tractor machinery, we can describe gravity and bosonic theories in a manifestly local 
unit invariant way. For example, the action for  cosmological Einstein gravity coupled to a massive scalar field is given by~\footnote{The tractor metric~$\eta_{MN}=\begin{pmatrix}0&0&1\\[1mm]0&\eta_{mn}&0\\[1mm]1&0&0\end{pmatrix}$
is the~$SO(d,2)$ invariant one and is also a  weight zero, rank two tractor. We denote the inner product of two tractors with the tractor metric by a dot, so $A\cdot B\equiv A^M \eta_{MN}B^N$ (not to be confused with a period for the dot product of four vectors $a.b=a^\mu b_\mu$).}
\begin{equation}\label{cosmological-scalar}\begin{split}
S=-\frac{d(d-1)}{2}\int \frac{\sqrt{-g}}{\sigma^d} \, &\Big[I \cdot  I +\frac{2\lambda}{d(d-1)}\Big]\\&- \frac12 \int \frac{\sqrt{-g}}{\sigma^{d+2w-1}} \, \varphi\, 
\, I\cdot D\,  \varphi\, .
\end{split}
\end{equation}
Here we have introduced the dimensionless combination of the cosmological constant and Newton's constant
$\lambda=\kappa^{\frac{4}{d-2}}\Lambda$ as well as the scale tractor
\be\nn
I^M=\frac1d D^M\sigma =\sigma\, \begin{pmatrix}1\\[3mm] b^m\\[1mm] -\frac1d\Big[\Rho+\nabla.b+b.b\Big]\end{pmatrix}\, ,\qquad b_\mu=\sigma^{-1}\partial_\mu\sigma\, .
\ee
The scale tractor controls the coupling of matter to scale. Moreover it is tractor parallel 
$$
{\cal D}_\mu I^M=0\, ,
$$
exactly when the metric~$g_{\mu\nu}$ is conformally Einstein. Its length~$I\cdot I$ is therefore 
constant in this case. In general however, the length of the scale tractor is Weyl invariant
but spacetime dependent.

The action~$S$ above is Weyl invariant precisely when the scalar~$\varphi$ has weight~$w$
\be\label{scaleq}
S[g_{\mu\nu},\sigma,\varphi]= S[\Omega^2g_{\mu\nu},\Omega\sigma,\Omega^w\varphi]\, .
\ee
Moreover the scalar equation of motion~$I\cdot  D\,  \varphi=0$ explicitly says
\be\nn
\Big[-\sigma^{2}g^{\mu\nu}\wt{ \nabla}_\mu\wt\nabla_\nu+ m^2_{\rm grav}\Big]\varphi=0\, .
\ee
The first term is the Weyl compensated scalar wave equation
where the $\wt \nabla_\mu$ denotes the Weyl compensated covariant derivative (it acts on scalars as $\wt\nabla_\mu = \nabla_\mu -w b_\mu$).
The second term in the scalar equation of motion is a ``gravitational mass'' term
\be\label{mass Weyl}
m^2_{\rm grav}=w(d+w-1)\,  I\cdot  I~.
\ee
If the gravitational background is taken on-shell and back-reaction is ignored, then the square of the scale tractor $I\cdot I$ is constant and the above ``mass-Weyl weight relationship'' relates masses to weights of tractors.
In fact, it generalizes to all higher spins, and allows mass to be reinterpreted as weight~\cite{Gover:2008sw}. However, since
the constant of proportionality is of order of the cosmological constant this introduces an unnatural tuning of weights
to masses. We remedy this situation in the next sections by accounting for back reaction.

\section{Tractor Stress Tensors}
\label{TST}

In the previous section, we saw that tractors were the natural language in which to formulate physics
in a way manifestly independent of local choices of unit systems.
Once matter systems are coupled to gravity, the natural way to present the field equations is a tractor generalization of Einstein's equations, along the lines of
\begin{equation}\label{EINST}
 \frak{G}^{MN}+\lambda\ \eta^{MN}=\frak{T}^{MN}\;,
\end{equation}
where $\frak{G}^{MN}$ and $\frak{T}^{MN}$ are symmetric tractor tensors built from the usual stress energy tensor and Einstein tensor. Our first step is to  construct a ``tractor stress tensor'' $\frak{T}^{MN}$, transforming correctly under Weyl rescalings. A locally unit invariant description of the matter sector of a theory yields, in general,  a Weyl invariant action for the matter fields $\phi_i$:
\begin{equation}\nn
S_{\rm Matter} [g,\sigma,\phi_i]=S_{\rm Matter}[\Omega^2g, \Omega\sigma, {\cal F}_i(\Omega,\phi_i)]\;.
\end{equation}
In this formula the transformation of the matter fields $\phi_i\mapsto {\cal F}_i(\Omega,\phi_i)$ is denoted
by the some function ${\cal F}_i$, which can often be easily determined using tractor technology along the lines
of~\cite{Gover:2008sw}. For the following arguments, its precise form is irrelevant.

The standard stress energy tensor of general relativity is given by
$$
T_{\mu\nu}(g,\sigma, \phi_i)=-\frac{2}{\sqrt{-g}}\frac{\delta S_{\rm Matter}}{\delta g^{\mu\nu}}\, .
$$ 
Clearly,  under Weyl transformations, it obeys 
$$T_{\mu\nu}\mapsto\Omega^{2-d}T_{\mu\nu}\, .$$ Thus, the most natural ingredient of our construction will be the Weyl invariant stress tensor:
\begin{equation}
\frak{T}^{mn}(g,\sigma,\phi_i)=-\frac{2}{\sqrt{-g}}\sigma^d e^{\mu m}e^{\nu n}\frac{\delta S_{\rm Matter}}{\delta g^{\mu\nu}}\;.
\end{equation}
It is worthwhile noting that the scale $\sigma$ enters in $\frak{T}^{mn}$ in a non trivial way, and that the usual stress tensor (apart from some powers of $\kappa$) is recovered once one chooses the canonical constant scale $\sigma=\kappa^{\frac{2}{d-2}}$. 

The tractor stress tensor we want to construct should contain  various physical objects related to $\frak{T}^{mn}$,  its trace $\frak{T}=\frak{T}^m_m$, divergence $\nabla. {\frak T}^m$ and so on. The only requirement needed to completely fix  $\frak{T}^{MN}$ is Weyl covariance, namely
\begin{equation}
\frak{T}^{MN}\mapsto U^M{}_R\,U^N{}_S\,\frak{T}^{RS}\;,
\end{equation}
where the tractor gauge transformation is given by~(\ref{tractor gauge transformation}). Imposing this transformation law, we are able to package $\frak{T}^{mn}$, its trace and derivatives, into the following tractor stress tensor:
\begin{equation}\label{tractor T MN}
{\frak T}^{MN}=\begin{pmatrix}
0 & 0 & \frac{1}{d}{\frak T} \\[2mm]
0 & {\frak T}^{mn} &-\frac{1}{d}\nabla. {\frak T}^m \\[2mm]
\ \frac{1}{d}{\frak T}\ \   & -\frac{1}{d}\nabla. {\frak T}^m & {\frak T}^{--}
\end{pmatrix}\;,
\end{equation}
with 
$$
{\frak T}^{--}=\frac{1}{d(d-1)}\Big[\nabla.\nabla. {\frak T}+\frac{\Delta {\frak T}}{d-2}+\Rho_{mn}\Big(d\, {\frak T}^{mn}-\eta^{mn}{\frak T}\Big)\Big]\, .
$$
These couplings of the stress tensor to the curvature are precisely those required to ensure that ${\frak T}^{MN}$
is a rank~2, weight zero, symmetric tractor tensor.

Now, let us turn to the gravity side. Of course, exactly the same procedure as for the matter sector applies.
One starts by making any choice of gravitational theory with action principle built from the metric. 
Then, by introducing the scale~$\sigma$, one rewrites the theory in a locally invariant way:
$$
S_{\rm Gravity}[g,\sigma]=S_{\rm Gravity}[\Omega^2g,\Omega\sigma]\, .
$$
Then one computes the tensor ${\frak T}^{mn}_{\rm Gravity}$ by exactly the same procedure as above with
$S_{\rm Gravity}$ in the place of $S_{\rm Matter}$ and forms the tractor tensor ${\frak T}^{MN}_{\rm Gravity}$.
The equations of motion are then simply ${\frak T}^{MN}_{\rm Gravity} + {\frak T}^{MN} = 0$. Specializing to the case
of the Einstein-Hilbert action, the gravity stress tensor is minus the Einstein tensor, and the same applies
to its tractor generalization in~\eqn{EINST}. An explicit computation (by varying the action $S=-\frac{d(d-1)}2\int \frac{\sqrt{-g}}{\sigma^d} I\cdot I$) yields
\bea\nn
{\frak G}^{mn}&=&\sigma^{2}\ G^{mn}(\sigma^{-2}g)\\[1mm]&=&
\sigma^2\ \Big\{G^{mn}(g)+(d-2)\Big[\nabla^m b^n+b^mb^n-\eta^{mn}\big(\nabla. b-\frac{d-3}2\ b.b\big)\Big]\Big\}\;.\nn
\eea
where $G_{\mu\nu}$ is the standard Einstein tensor and the first line, of course, is the invariant tensor obtained from it by Weyl compensating. The tractor Einstein tensor is therefore
$$
{\frak G}^{MN}=\begin{pmatrix}
0 & 0 & \frac{1}{d}\frak G \\[2mm]
0 & \frak G^{mn} & -\frac 1d \nabla.\frak G ^m\\[2mm]
\frac{1}{d}\frak G & -\frac 1d \nabla.\frak G ^m & \frak G^{--}
\end{pmatrix}\;,
$$
with $\frak G^{--}$ given by the analogous formula to $\frak T^{--}$ in~\eqn{tractor T MN}.
At the canonical choice of scale $\sigma=\kappa^{\frac2{d-2}}$ we find
\begin{equation}
\frak G^{MN}(g,\kappa^{\frac{2}{d-2}})=\kappa^{\frac{4}{d-2}}\begin{pmatrix}
0 & 0 & \frac{1}{d}\ G \\[3mm]
0 & G^{mn} & 0\\[2mm]
\ \frac{1}{d}\, G \ \ \ \ \ \ \ & 0 & \ \frak G^{--}(g,\kappa^{\frac{2}{d-2}})
\end{pmatrix}\;,
\end{equation}
where 
$$
\frak G^{--}(g,\kappa^{\frac{2}{d-2}})=\frac{d-2}{d-1}\ \Rho_{mn}(\Rho^{mn}-\frac1d\eta^{mn}\Rho)-\frac1d\Delta\Rho\, .
$$
Vanishing of $\frak G^{MN}(g,\kappa^{\frac{2}{d-2}})$ is exactly Einstein's equations in vacuum. For general
choices of local unit systems, $\frak G^{MN}=0$, implies that the metric is conformally Ricci flat. Adding the
cosmological term, so that $\frak G^{MN}+\lambda \eta^{MN}=0$, says the metric is conformally Einstein.

It is worth mentioning that the conformally Einstein condition can also be simply expressed as
$$
D^M I^N = 0\, .
$$
Indeed, this relationship holds because the tractor Einstein tensor can be written~as
\be\label{GID}\begin{split} 
\frak G^{MN} = \sigma D^M I^N &-X^{(M} D^{N)}
I\cdot I+\frac{(d-1)(d-2)}{2}\eta^{MN} I\cdot I\\&-\frac1{(d-1)(d-2)^2}X^M X^N
\Big( D_R I_S \ D^R I^S \Big)~, 
\end{split}\ee
where $X^M =\left({\tiny\begin{array}{c} 0\\0\\1 \end{array}}\right)$ is the
Weyl-invariant weight one ``canonical tractor.''

Finally, we have so far discussed variations with respect to the metric, but in a locally 
unit invariant description of physics, one generally incorporates the scale field~$\sigma$
which must also be varied.
However, for any locally unit invariant action $S[g,\sigma]=S[\Omega^2g,\Omega\sigma]$
we have $\delta S[g,\sigma]/\delta\sigma=\delta S[\sigma^{-2}g,1]/\delta\sigma=-\frac{\sqrt{-g}}\sigma T_{\mu}^{\mu}$.
Examining~\eqn{tractor T MN}, we see therefore that the~$\sigma$ field equation is not a new relation but
simply implies\footnote{One might be tempted to think that the Weyl invariance of $S[g,\sigma]$ would imply
vanishing of the stress energy trace as identity ({\it i.e.}, off-shell). However, since this invariance is achieved via compensating, in fact no new identity follows.} $\frak T^M_{M\, \rm Gravity}$ $+$ $\frak T^M_M=0$. 
Specializing to cosmological Einstein gravity coupled to matter, this field equation reads
\be\label{trace}
\frak G^{M}_M+(d+2)\lambda = \frak T^M_M\, ,
\ee
and will play a special {\it r\^ole} in our treatment of back-reaction in the next section.

\section{Tractor Back-reaction}
\label{TBR}

In section~\ref{tractors}, we showed how local unit invariance implied that mass could be replaced with 
the more fundamental geometric notion of weight. (To be precise the weight of a tractor field.) This has
the advantage that masses no longer need be defined in terms of quadratic Casimirs of the spacetime
isometries, but rather in terms of the geometric field content of the theory. This is extremely appealing, 
because no recourse to backgrounds with special isometries is needed. The drawback of this approach
is the unnaturalness of the  relationship between masses and weights~\eqn{mass Weyl}, relying on the cosmological 
constant to set fundamental mass scales. We now show how back-reaction immediately solves this problem.

Consider the simplest case of a massive scalar field coupled to cosmological Einstein gravity (whose tractor
version of the standard action principle is given in~(\ref{cosmological-scalar})). The full equations of motion,
including back-reaction are
\bea\nn
I\cdot D\  \varphi\ \ &=&\ 0\, ,\\[2mm]
{\frak G}^{MN}+\lambda \eta^{MN}&=&{\frak T}^{MN}\, .\nn
\eea
In the lowest approximation, ignoring back-reaction, the solution of the second equation is an Einstein
metric with constant scalar curvature, or in tractor language, constant length scale tractor, $I\cdot I=$ constant.
As explained in section~\ref{tractors}, the first equation describes a standard massive scalar field in 
an Einstein background with constant mass determined by the weight $w$ of~$\varphi$ by~\eqn{mass Weyl}.
Once we include back-reaction, we are faced with the above coupled set of equations, whose novel feature is
that the mass term for the scalar field is proportional to the spacetime dependent (but Weyl invariant) quantity $I\cdot I$.
Explicitly (transcribing~\eqn{scaleq}) this reads
\be\label{eigenval}
\sigma^2g^{\mu\nu} \wt \nabla_\mu\wt \nabla_\nu\,  \varphi = w(d+w-1) I\cdot I\, \varphi\, .
\ee
However, in vacua, the length of the scale tractor is set by the trace of the tractor Einstein tensor (or in commonplace language,
the trace of the Einstein tensor is proportional to the scalar curvature). We must therefore correct this relationship 
to include backreaction.
To be precise, tracing~\eqn{GID} with the 
tractor metric and using~(\ref{trace},\ref{tractor T MN}) we obtain
\be\nn
\frac12(d-1)(d-2)(d+2) I\cdot I + (d+2)\lambda=\frak T^M_M\, . 
\ee
Hence
\be\nn
\sigma^{2}g^{\mu\nu}\wt \nabla_\mu\wt\nabla_\nu\,  \varphi = -\frac{2w(d+w-1)}{(d-1)(d-2)} \, \Big[\, \lambda\, -\, \frac1{d}\frak T\, \Big] \;\varphi\, .
\ee
In the canonical choice of scale we have

\begin{center}
\shabox{
$
\Delta\varphi = -\frac{2w(d+w-1)}{(d-1)(d-2)}\, \ \Big[\Lambda-\frac{\kappa^{-\frac{4w}{d-2}}}{d} \, T(g,\varphi)\Big]\ \varphi\, ,
$}
\vspace{-7mm}	
\end{center}
\be\label{central}\ee 
where $T(g,\varphi)$ is the trace of the standard stress tensor for this system.
Something extremely important has happened in this equation that makes it the  central one of this paper:
In a theory with a constant mass term, the right hand side of~\eqn{eigenval} would not have been modified as it
is above by back-reaction. The tractor theory, on the other hand, coincides with the standard theory in the absence of
back-reaction, but cleverly replaces the cosmological constant $\Lambda$ by
\be\nn
\Lambda\longrightarrow \Lambda -\frac{\kappa^{-\frac{4w}{d-2}}}{d} \, T\, \,
\ee
in the back-reacting mass term. 

Now let us recompute the mass-Weyl weight relationship to include back-reaction. For that, 
there are two routes we could take. In fact this is a critical juncture of this paper.
The first route is to insert the trace of the matter stress tensor computed from the classical matter action.
There is no way for this computation to know about scales where particle physics takes place, so it is unlikely
that it can solve our naturalness problem. Nonetheless, the classical computation does generate a rather interesting
non-linear potential for the scalar field which is described in the appendix (a study of the effects
of the $\varphi$ dependence of $T(g,\varphi)$).  The second---more exciting---route is a phenomenological 
approach taking into account quantum effects. Our observation is simple, namely, once we introduce a particle physics
scale into the problem (either by cutting off divergences at some characteristic scale, or augmenting the toy-scalar
field model with standard model fields), the expectation of the trace of the stress tensor will be of the same scale.
To investigate the leading order quantum effects of integrating out the gravity modes, therefore, we must
replace
the trace of the stress tensor with its vacuum expectation value and
find
\be\nn
m^2_{\rm Back-reaction}= -\frac{2w(d+w-1)}{(d-1)(d-2)}\, \Big[\Lambda - \frac{\kappa^{-\frac{4w}{d-2}}}{d}\, \langle T \rangle\Big]\, .
\ee
This immediately solves our naturalness problem! No longer does the cosmological constant alone set the scale
of the mass term, but instead it appears in combination with a particle physics scale object -- the trace of the stress tensor. We emphasize that this should not be viewed as a solution to the cosmological constant problem, but it is
an example of a solution to a problem caused by the cosmological constant.

 We close this section by noting that the above
mechanism is germane to particles of arbitrary spin. The idea is no different to above.
In the work of~\cite{Gover:2008sw}, it was shown that massive, massless and partially massless~\cite{Deser:1983tm,Higuchi:1986py,Deser:2001pe}
fields of arbitrary spin~$s$ could be written in a manifestly unit covariant
way using symmetric, weight $w$ tractors $\varphi^{M_1\ldots M_s}$ of rank~$s$. The equations of motion are simply
\bea
I\cdot D \varphi^{M_1\ldots M_s} &=&0\, ,\nn\\
D\cdot \varphi^{M_2\ldots M_s} \ &=&0\, ,\nn\\
I\cdot \varphi^{M_2\ldots M_s} \ \ &=&0\, ,\nn\\
\varphi_M^{MM_3\ldots M_s}\ \ \  &=&0\, .\label{shell}
\eea
The last three equations are constraints (or generalized Feynman-type gauges for massless theories)
ensuring the correct propagating degrees of freedom. The first equation implies a Klein--Gordon equation for the
physical modes $\varphi^{\mu_1\ldots\mu_s}$ (subject to $\nabla_\mu\varphi^{\mu\mu_2\ldots\mu_s}=0=
\varphi_\mu^{\mu\mu_3\ldots\mu_s}$)
\be\nn
\sigma^{2}g^{\mu\nu}\wt\nabla_\mu\wt\nabla_\nu \varphi^{\mu_1\ldots\mu_s}= [w(w+d-1)-s]\, I\cdot I\, \varphi^{\mu_1\ldots\mu_s}\, .
\ee
This brings us to the starting point~\eqn{eigenval} for our discussion of the back-reacting massive scalar field. 
From here the discussion proceeds exactly as above. Moreover, an analogous discussion for Fermi fields can be carried
out, since a Weitzenbock type identity for those models yields again an equation of the above Klein--Gordon type (see~\cite{Gover:2008sw}
for details). Finally, we make note that our analysis applies equally well to massless and partially massless theories
since they are described also by~\eqn{shell}: Strictly massless theories are obtained by setting the weight $w$ to $w=s-2$
while depth $t$ partially massless theories (which enjoy higher derivative gauge symmetries of order~$t$) occur at $w=s-t-1$
for $2\leq t\leq s$.

It is highly gratifying to see the first corrections due to back-reaction of the unit covariant description of physical
systems solve a naturalness problem introduced by the cosmological constant. It is natural to speculate, that even
deeper insights into how to handle the vast differences in scale between particle and gravitational physics might follow
from this viewpoint. To make inroads into this historically very difficult problem, we must surely tackle the problem 
of quantization head-on in  our approach. Therefore we close this paper with some more speculative remarks about
renormalization and scale anomalies.

\section{Holographic Anomalies}

So far we have concerned ourselves with an essentially classical, but manifestly unit invariant, analysis of field theories.
However, since Weyl and scale invariances are typically anomalous, it is extremely interesting to apply our
ideas to quantum field theories. In particular we close this paper by considering renormalization theory, one of the key ingredients of quantum field theory. This discussion will necessarily be both brief and somewhat speculative, 
but we cannot resist mentioning some very exciting future directions for the tractor program. 

\begin{figure}
\begin{center}
\includegraphics[width=7cm]{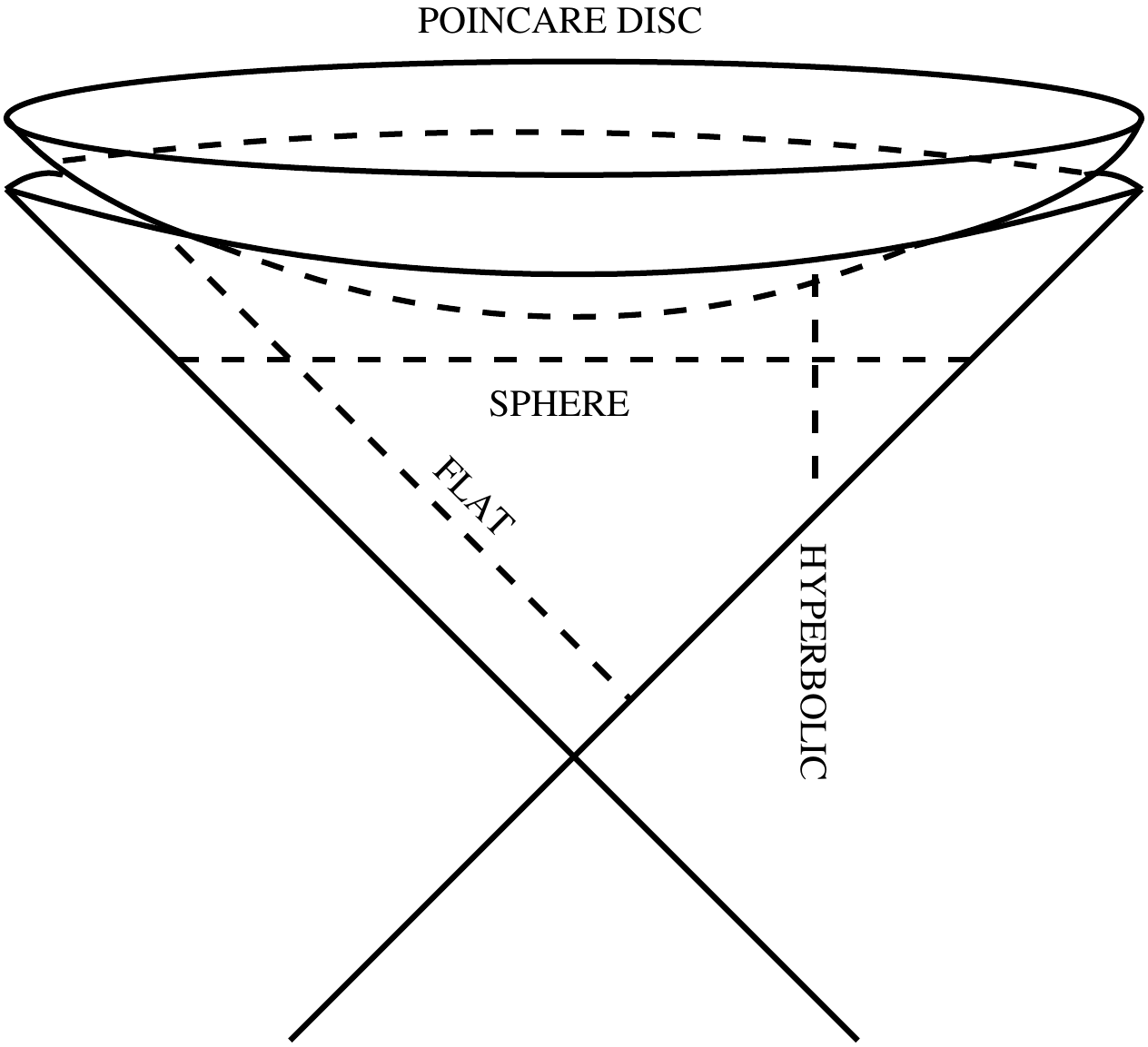}
\end{center}
\caption{Cutting a lightcone with  a horizontal plane yields a sphere (with canonical metric induced from the
flat, Lorentzian ambient space). A diagonal plane gives flat space, and hyperbolic space comes from a vertical one.
The Poincar\'e model for conformal geometries can also be embedded in this picture. \label{coneanddisc}}
\end{figure}

Among the more stunning implications of Maldacena's AdS/CFT correspondence~\cite{Maldacena:1997re} is the formulation of renormalization
group flows of boundary quantum field theories in terms of geometry of a bulk theory of one higher dimension~\cite{de Boer:1999xf}.
In particular, boundary Weyl anomalies appear as coefficients of logarithms of the radial coordinate when attempting  to perform expansion of the bulk metric away from the boundary characterized by a conformal class of
boundary metrics~\cite{Henningson:1998gx}
In the mathematics literature this is known as a Fefferman--Graham expansion~\cite{FG} and
the bulk metric (the AdS side of the physics correspondence) is referred to as a Poincar\'e metric (in reference
to the Poincar\'e disc as the flat, Riemannian signature, model for these geometries). However, as we mentioned before, a persistent theme
in studies of Weyl invariance (dating back to Dirac~\cite{Dirac}) is the importance of a $(d+2)$-dimensional ``ambient'' spacetime. For conformally flat geometries this can be understood from the picture in figure~\ref{coneanddisc}
where (in Riemannian signature) it is shown how to obtain various conformally equivalent metrics from slicings of a $(d+2)$-dimensional lightcone. (In fact the entire conformally flat equivalence class of metrics can be obtained this way.)
Moreover, the embedding of Poincar\'e disc as a hyperboloid in $(d+2)$-dimensions is depicted in order to indicate 
the identification between its boundary manifold and the conformal manifold described by lightlike rays. In fact this ambient description of conformal manifolds extends to general conformal classes of metrics (as already described in the original work of Fefferman and Graham~\cite{FG}) and also yields an ambient description of tractor calculus. The key idea being to endow the ambient space with a $(d+2)$-dimensional Ricci flat metric $g_{MN}$ admitting a hypersurface orthogonal homothety characterized by
\be\nn
g_{MN}=\nabla_M X_N\, .
\ee
There is no symmetrization on the right-hand side of this equation and $X_M$ stands for the covariant components 
of homothetic vector field whose square vanishes on the curved generalization of a $(d+2)$-dimensional lightcone.
This picture suggests a Ricci flat/CFT correspondence in which renormalization group flows become geometric in a space of two higher dimensions. Of course this idea is easier stated than carried out. In particular
an exact correspondence between Ricci flat ``bulk'' data and boundary correlators is needed. Moreover, 
one would in principle need to control the quantization of the Ricci-flat bulk. However, even though
there is the complication of the extra constraint that the metric possess a hypersurface orthogonal homethety $X^M$ in order to define the ambient lightcone, the absence of a bulk cosmological constant may make the problem
more amenable to a String Theory solution.
 
There are two further approaches to quantization suggested by the ambient space picture. The first is closely 
related to the original AdS/CFT approach, but formulated in a way that manifests local scale invariance:
An ambient space  scalar~$\sigma$ subject to a unit weight condition $X^M\nabla_M\,  \sigma=\sigma$, 
amounts to a canonical choice of metric in the $d$-dimensional conformal class of metrics (the field $\sigma$ is
the ambient version of the scale $\sigma(x)$). In particular, choosing $\sigma$ so that the scale tractor is
parallel and non-null yields an Einstein manifold with non-zero cosmological constant. Then identifying this manifold 
with the bulk theory one can try to construct a correspondence with a $(d-1)$-dimensional boundary conformal 
field theory. Since the canonical bulk metric appears at constant values of the scale $\sigma$, it makes sense to
express bulk quantities as power series in $\sigma$. Again, one can study obstructions in this expansion to  
obtain boundary Weyl anomalies~\cite{GW}.

The final ambient space approach is more radical than those above. The key idea is to realize that the 
$(d+2)$-dimensional lightcone depicted in~\ref{coneanddisc} can be identified with the space of on-shell 
excitations of massless systems. This is achieved by viewing the ambient space as a momentum space.
This identification was made completely explicit for a six-dimensional scalar field in the $d$-dimensional
conformally flat setting in~\cite{Gover:2009vc}. In that work, it was also shown how to generalize the momentum space side of
this identification to curved space and general conformal classes of metrics. A welcome byproduct of this
approach was a novel derivation of tractor operators from generators of the conformal group $SO(d+1,3)$ of the ambient space itself. This duality suggests a far more general one in which four dimensional physics is embedded
in the momentum space of some six-dimensional theory whose quantization controls  that of our underlying
four-dimensional theory.

\section*{Acknowledgments}
It is a pleasure to thank Itzhak Bars, Fiorenzo Bastianelli, Nicholas
Boulanger, Claudia de Rham, Rod Gover, Maxim Grigoriev, Augusto
Sagnotti, Abrar Shaukat, Per Sundell, Misha Vasiliev and Roberto
Zucchini for useful discussions. A.W. thanks the INFN, Scuola Normale
Pisa and the University of Bologna for hospitality. The work of
R.B. and O.C. is partly supported by the Italian MIUR-PRIN contract 20075ATT78.

\appendix

\section{Higher Order Classical Back-reaction}

In this appendix we compute the higher order classical corrections to the equation of motion for a massive
scalar with a unit-covariant, scale tractor coupling to cosmological
Einstein gravity. Starting from 
the central equation~\eqn{central}, we must first compute the trace of
the stress energy tensor. We find 
$$
T(g,\varphi)=\frac{(d+2w)(d+2w-2)}{2d}\,\Big[-\pa_\mu\varphi \, g^{\mu\nu}\partial_\nu\varphi+\frac{2w(d+w-1)}{d}\Rho\varphi^2
  \Big]~.  
$$
Firstly notice that the latter vanishes for
$w=1-\frac{d}2,-\frac{d}2$. These weights correspond to having a
scalar field conformally coupled to gravity $\int \varphi\big(\Delta
-\frac{(d-2)}2 \Rho\big)\varphi$. The first special weight is the engineering dimension
for the conformal scalar field in $d$-dimensions, whereas the second simply
amounts to a trivial field redefinition of the conformal scalar $\varphi\to\varphi/M_{\rm Pl}$.  
For generic weights the trace of the Schouten tensor~$\Rho$ can be computed in terms
of~$\varphi$ from~\eqn{trace} which yields
\begin{equation}
T(g,\varphi)=\frac{(d+2w)(d+2w-2)}{2d}\,\frac{c_1\Lambda\varphi^2-\pa_\mu\varphi\,  g^{\mu\nu}\partial_\nu\varphi}{1+c_2\kappa^{-\frac{4w}{d-2}}\varphi^2}\;,
\end{equation}
where
$
c_1= \frac{2w(d+w-1)}{(d-1)(d-2)}$ and $c_2=\frac{w(d+2w)(d+w-1)(d+2w-2)}{d^2(d-1)(d-2)}
$. Note that both $c_1$ and $c_2$ are invariant under $w\rightarrow 1-w-d$.

\begin{figure}
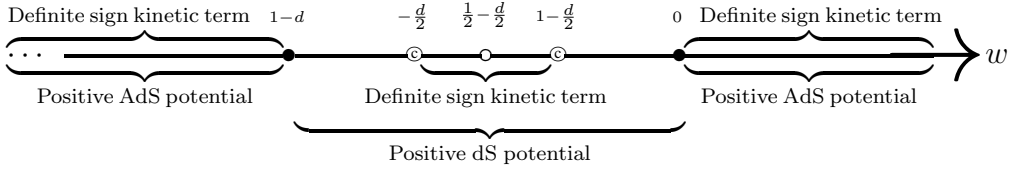

\begin{center}
$$
\underbrace{\overbrace{\cdots\phantom{\circ}\rule[.7mm]{2.9cm}{0.5mm}}^{\rm Definite\: sign\: kinetic\: term\ \  \ \ }}_{\rm Positive\: AdS\: potential}
\hspace{-3mm}
	\stackrel{{}^{{}^{{}^{1-d}}}}{\bullet}
\hspace{-3mm}
\underbrace{
\rule[.7mm]{1.5cm}{0.5mm}
\hspace{-2.6mm}
	\stackrel{{}^{{}^{{}^{-\frac{d}2}}}}{\raisebox{.6mm}{\tiny\copyright}}
\hspace{-2.7mm}
\underbrace{
\rule[.7mm]{.8cm}{0.5mm}
\hspace{-4mm}
	\stackrel{{}^{{}^{{}^{\frac12-\frac{d}2}}}}{\circ}
\hspace{-4mm}
\rule[.7mm]{.8cm}{0.5mm}
}_{\hspace{-.8cm}\raisebox{2mm}{$\scriptstyle\rm Definite\: sign\: kinetic\: term$} \hspace{-.8cm}}
\hspace{-3.2mm}
	\stackrel{{}^{{}^{{}^{1-\frac{d}2}}}}{\raisebox{.6mm}{\tiny\copyright}}
\hspace{-3.3mm}
\rule[.7mm]{1.6cm}{0.5mm}
}_{{\rm Positive\;  dS\;  potential}}
\hspace{-3mm}
	\stackrel{{}^{{}^{{}^{0}}}}{\bullet}
\hspace{-2.7mm}
\overbrace{\phantom{\circ}\!\!\!\!
\underbrace{
\rule[.7mm]{3.3cm}{0.5mm}}_{\rm Positive\: AdS\: potential}}^{\ \ \ \rm Definite\: sign\: kinetic\: term}\!\!\!\!\!\!\!\!\!\!\!\!
\raisebox{-2.3mm}{\scalebox{3.2}{$\rightarrow$}}w
$$
\end{center}
\caption{Behavior of kinetic and potential terms as a function of
  $w$. The circle~$\circ$ denotes the value of $w$ corresponding to a
  mass saturating the Breitenlohner--Freedman bound. The value
  corresponding to  a conformally improved scalar field is denoted by
  ${\copyright}$.
  \label{ranges}} 
\end{figure}

Putting classical gravity on-shell ({\it i.e.} tree-level graviton exchange\footnote{Actually we are putting $R(g)$ 
on-shell, but the as-yet unknown, dynamical, metric $g_{\mu\nu}$ still resides in 
the terms $\pa_\mu\varphi \, g^{\mu\nu}\partial_\nu\varphi$ and $\Delta\varphi$.})  seems to generate rather complicated non-linear scalar self interactions since the scalar equation~\eqn{central} now takes the explicit form:
\begin{equation}
\Delta\vphi+c_1\Lambda\vphi+c_2\kappa^{-\frac{4w}{d-2}}\,\frac{\pa_\mu\varphi \, g^{\mu\nu}\partial_\nu\varphi}{1+c_2\kappa^{-\frac{4w}{d-2}}\,\vphi^2}-c_1c_2\kappa^{-\frac{4w}{d-2}}\,\frac{\Lambda\,\vphi^3}{1+c_2\kappa^{-\frac{4w}{d-2}}\,\vphi^2}=0\, .
\end{equation}
However this equation in fact  follows 
 by varying the  simple non-linear sigma model action:
\begin{equation}\label{sigma model}
S=\int\sqrt{-g}\Big[-\frac12\,G(\vphi)\, \pa_\mu\varphi \, g^{\mu\nu}\partial_\nu\varphi-U(\vphi)\Big]\;, 
\end{equation}
with sigma model metric and potential given by
$$
G(\vphi)=1+c_2\kappa^{-\frac{4w}{d-2}}\,\vphi^2\;,\quad U(\vphi)=-\frac{c_1}{2}\,\Lambda\,\vphi^2\;.
$$Classical back-reaction has thus generated scalar interactions that are encoded in the above lagrangian. By means of a field redefinition \be\label{redef}G(\vphi)d\vphi^2=d\chi^2\, ,\ee one can generically put \eqn{sigma model} in the form of a linear sigma model with a 
non trivial potential,~\emph{i.e.}
$$
S=\int\sqrt{-g}\Big[-\frac12\,\pa_\mu\chi\,g^{\mu\nu}\pa_\nu\chi+\frac{c_1}{2}\Lambda\vphi^2(\chi)\Big]\;.
$$
The above field redefinition comes with a {\it caveat}: the kinetic term must have definite sign for all field configurations.
This is true only when $c_2$ is positive which only holds for certain values of~$w$. Another interesting feature of
the model is that the potential is only positive when $-\Lambda c_1>0$ which again holds only for certain ranges of~$w$.
These ranges are depicted in figure~\ref{ranges}. We do not perform a detailed phenomenological analysis
of this model here, but note that it is rather interesting to see the  rich structure introduced by this simple toy model.

\end{document}